\documentclass{article}
\usepackage{emulateapj,graphics,onecolfloat}

\font\tfa=cmr10 at 8.00pt
\tfa

\begin{document}
\twocolumn[
\title {Surveys of Galaxy Clusters with the Sunyaev Zel'dovich Effect}
\author {A.E. Schulz$^{1}$, Martin White$^{2}$}
\affil{$^{1}$Department of Physics, Harvard University,
Cambridge, MA 02138}
\affil{$^{2}$Departments of Physics and Astronomy, University of California,
Berkeley, CA 94720}

\begin{abstract}
\noindent
\rightskip=0pt
We have created mock Sunyaev-Zel'dovich effect (SZE) surveys of galaxy
clusters using high resolution N-body simulations.  To the pure surveys
we add `noise' contributions appropriate to instrument and primary CMB
anisotropies.
Applying various cluster finding strategies to these mock surveys we generate
catalogues which can be compared to the known positions and masses of the
clusters in the simulations.
We thus show that the completeness and efficiency that can be achieved depend
strongly on the frequency coverage, noise and beam characteristics of the
instruments, as well as on the candidate threshold.
We study the effects of matched filtering techniques on completeness, and bias.
We suggest a gentler filtering method than matched filtering in single
frequency analyses.  We summarize the complications that arise when analyzing
the SZE signal at a single frequency, and assess the limitations of such an
analysis.  Our results suggest that some sophistication is required when
searching for `clusters' within an SZE map.
\end{abstract}

]

\rightskip=0pt

\section{Introduction}

Observations of the number density of clusters of galaxies will play an
increasingly important role in determining the composition of the energy
density in the universe as data from the myriad of upcoming cluster surveys
accumulates.  Cluster surveys result in constraints orthogonal in parameter
space to those obtained from other cosmological observations, such as the
Cosmic Microwave Background (CMB) anisotropies and supernova searches, because
the cluster abundance depends significantly on the linear growth function.
For this reason, clusters can also be used to probe the nature and evolution
of the dark energy in the universe.
Since clusters are the most recently formed gravitationally bound objects in
the universe, the evolution of their number density sensitively probes the
critical redshift range $0<z<2$, a range over over which the dark energy has
come to dominate the total energy density.
Clusters are convenient in that they are very bright, and rare enough to make
counting them tractable.

A wide array of survey techniques is being used to conduct searches for
clusters, making use of optical and X-ray emissions from clusters, weak
lensing distortions, and the Sunyaev Zel'dovich effect
(SZE; see Table~\ref{tab:expts}).
The SZE is a particularly promising approach for finding galaxy clusters
because the signal is relatively independent of the cluster's distance from us.
This implies that, at least in principle, the selection function for such
surveys is very well known.  This is crucial if we are to use the cluster
catalogues to measure the evolution of the number density of clusters.

\begin{table*}
\begin{center}
\begin{tabular}{lcccl}
Name	& Type	& Freq (GHz)	& Resoln	& Web site \\  \hline
Acbar	& Bolo	& 150-270	&	4	&
  http://cosmology.berkeley.edu/group/swlh/acbar/ \\
Bolocam	& Bolo	& 145		&      1	&
  http://www.astro.caltech.edu/$\sim$lgg/bolocam\_front.htm \\
CBI	& HEMT	& 30 &	4.5	& http://www.astro.caltech.edu/$\sim$tjp/CBI/ \\
SZA	& HEMT	& 30 & $\sim 1$	& http://astro.uchicago.edu/SZE/ \\
AMI	& HEMT	& 15 & $\sim 1$	& http://www.mrao.cam.ac.uk/telescopes/ami/ \\
Amiba	& HEMT	& 30 & $\sim 1$	& http://www.asiaa.sinica.edu.tw/amiba/ \\
APEX	& Bolo	&150 &	0.75	& http://bolo.berkeley.edu/apexsz/ \\
SPT	& Bolo	&150 &	1	& http://astro.uchicago.edu/spt/ \\
Planck	& Bolo	&30-850&5	& http://astro.estec.esa.nl/Planck/
\end{tabular}
\end{center}
\caption{Some upcoming Sunyaev-Zel'dovich experiments.  Type indicates the
nature of the receivers, HEMTs or Bolometers.  The frequency is given in GHz.
The resolution is an estimate of the beam size, in arcminutes, and for the
interferometers this estimate is quite approximate.
The last 6 experiments intend to undertake blank field SZ surveys.
More information on these experiments can be found at the listed Web sites.}
\label{tab:expts}
\end{table*}

Using the SZE does present certain difficulties.  The energy lost
by the CMB photons on their journey from the last scattering surface is an
integrated effect.  Hence the SZE signal suffers from projection effects from
other objects in the same line of sight as the cluster, and also yields no
information about the redshift of the cluster save its angular size on the sky.
Thus, to study evolution effects in the number density, a followup observation
is required to obtain the redshifts, and in rare cases distinguish between
two separate clusters that lie in the same line of sight.  The primary CMB
anisotropies are also a problem, having significant power on cluster-sized
angular scales.  Finally, since clusters of galaxies are not perfect, isolated
spheres of gas, the SZE signal obtained from a cluster of a given mass will
vary considerably depending on the particular line of sight through the
cluster.
These effects make it difficult to correlate the SZE signal with the actual
mass of the cluster causing it.  In this paper we make a preliminary
investigation of the power of SZE surveys in finding clusters, taking these
effects into account, and present the results in terms of the completeness
and efficiency of mass limited samples achievable using the SZE.

\section{Method}

We construct maps of the SZE effect at various frequencies using as input
a high-resolution N-body simulation of structure formation in a
$\Lambda$CDM cosmology.  In this way our method is similar to that of
Kay, Liddle \& Thomas~(\cite{KayLidTho}).
We make mock observations of these maps by adding signal from the primary CMB
anisotropies to the SZE maps, convolving with a beam window function,
and adding Gaussian random noise such as might be produced by the electronics
in a real observation.
We identify every cluster candidate in these mock observations using
a specified method and check it against the true 3-D positions of the
clusters in the same simulation.

We present our results in terms of the completeness and efficiency of the
method in finding clusters above a mass threshold.
Completeness is the ratio of the number of clusters we found using the mock
SZE observation to the total number of massive clusters in the field of view.
Out of the total number of cluster candidates that we identify in our SZE
maps, only some of them will actually be clusters with a mass above the
threshold of interest.
Efficiency will measure the ratio of clusters found to the total number of
candidates, and is a measure of the amount of contamination suffered when
using the SZE technique.
The survey efficiency will be important in planning follow-up observations
with other instruments.
Obviously an SZE survey will be useful for many things besides creating an
effectively mass selected sample, but it is such a sample which is the
easiest to compare with theories of structure formation.
It has also become commonplace to describe SZE surveys as ``effectively
mass limited'', and it is for these reasons that we focus on this metric here.

\subsection{The N-body simulation}

The starting point for constructing the maps is an accurate model of the
spatial distribution of mass along the past light-cone.  We obtain this
{}from an N-body simulation of $512^3$ particles in a (periodic) cube of
side $300h^{-1}$Mpc run with a TreePM-SPH code (see the appendix of
White \cite{MassFn}).
Since on the scales of relevance to us baryonic pressure is sub-dominant,
only collisionless dark matter is modeled allowing us to achieve a higher
dynamic range in the simulation.  This allows us to simulate a larger
volume, containing more of the rare rich clusters we are interested in,
at the expense of an ad hoc (but flexible) treatment of the baryonic physics.
The simulation is started at $z=60$ and evolved to the present with the full
phase space distribution dumped every $100h^{-1}$Mpc between redshifts $2>z>0$.
It is this range of redshifts which dominates the SZE signal on the angular
scales of interest to us, but by cutting off the integration at $z=2$ we will
underestimate the effect of confusion in our maps.
The gravitational softening used is of a spline form, with a
``Plummer-equivalent'' (comoving) softening length of $20 h^{-1}$kpc.
We have used a flat cosmology compatible with a host of current
observations; $\Omega_{\rm m}=0.3$, $\Omega_\Lambda=0.7$,
$\Omega_{\rm b}h^2=0.02$, $h=0.7$, $n=1$, and $\sigma_8=1$.
The transfer function was evaluated with the fitting function of
Eisenstein \& Hu \cite{EisHu}.
While a slightly lower $\sigma_8$ would better fit the inferred mass function
of rich clusters from X-ray surveys, the higher $\sigma_8$ provides an easier
match to the CBI deep field observations (Mason et al.~\cite{CBI}).
The mass resolution in the simulation is fine enough to identify galactic mass
halos, with non-interacting dark matter particles of mass
$1.7\times 10^{10} h^{-1} M_{\odot}$.
All of the relevant cluster-scale halos contain several thousand particles to
begin to resolve sub-structure.
The simulation was performed on 128 processors of the IBM-SP2 at NERSC,
took nearly 4000 time steps and approximately 100 wall clock hours
to complete.

To construct the long thin line of sight used to compute the net SZE,
we have stacked the intermediate stages of the simulation between redshifts
$2>z>0$.
In order to avoid multiply sampling the same large scale structures, each
$300 h^{-1}$Mpc box has been randomly re-oriented in one of the six possible
orientations, and has furthermore been shifted by a random amount,
perpendicular to the line-of-sight, making use of the periodic boundary
conditions.
There are three time dumps per box length.
Each $300 h^{-1}$Mpc volume in the stack is made up of three segments, each
segment evolved to a later epoch than the previous one by the time it takes
light to travel $100h^{-1}$Mpc.  We have chosen $100 h^{-1}$Mpc as the
sampling interval because it is large enough that edge effects are minimal,
yet fine enough that the line of sight integrals are well approximated by
sums of the (static) outputs.  
Because of the periodicity, we are free to choose any of the thirds as the
oldest, cyclically permuting the other two.
This approach preserves the continuity of large-scale structure over distances
of $300 h^{-1}$Mpc without compromising the resolution in time evolution.   

\subsection{Cluster catalog}

In order to compute the completeness and efficiency with which the mock SZE
survey can detect clusters, we need to know the true distribution of cluster
in the simulated fields.  To this end we construct a catalog of the 3-D
position, redshift, mass, velocity dispersion, and other useful quantities of
each identified halo above $10^{13}h^{-1}M_{\odot}$.
Halos are identified using a friends-of-friends algorithm
(Davis et al.~\cite{DEFW})
on each of the time dumps used in the line of sight integral.
The FoF algorithm partitions the particles into equivalence classes by linking
together all particle pairs separated by less than a distance $b$.
We use a linking length of $b=0.15$ times the mean inter-particle spacing,
which is smaller than the typical value of $b=0.2$ because it reduces the
number of instances in which two separate halos, connected by a filament
of significant overdensity, get accidentally classified as a single object. 
Such misclassifications were found to be a significant source of confusion
when computing completeness and efficiency
(see also White \& Kochanek~2002, Kochanek et al.~2003,
White, van Waerbeke \& Mackey~2002 for further discussion).
While the spherical overdensity algorithm, which is less susceptible to the
merging problem, could have been used, this algorithm identifies spherical
clusters which would introduce a different type of bias.
We use position of the potential minimum of the FoF group as the center of
the cluster because it is more robust than using the center of mass,
coinciding closely with the density maximum of the cluster for all but the
most anomalous clusters.  Centering on the potential minimum we computed, for
each halo, the mass, $M_{200}$, enclosed within a radius, $r_{200}$, interior
to which the density contrast was 200 times the critical density. 
There are 6887 halos in the $z=0$ 3D catalogue with
$M_{200}>10^{13}\,h^{-1}M_\odot$ of which 790 are more massive than
$10^{14}\,h^{-1}M_\odot$ and 9 have masses greater than
$10^{15}\,h^{-1}M_\odot$.  A typical simulated field will contain
two or three hundred halos more massive than $10^{14}\,h^{-1}M_\odot$.

\subsection{Compton-$y$ maps}

\begin{figure}
\begin{center}
\resizebox{3.5in}{!}{\includegraphics{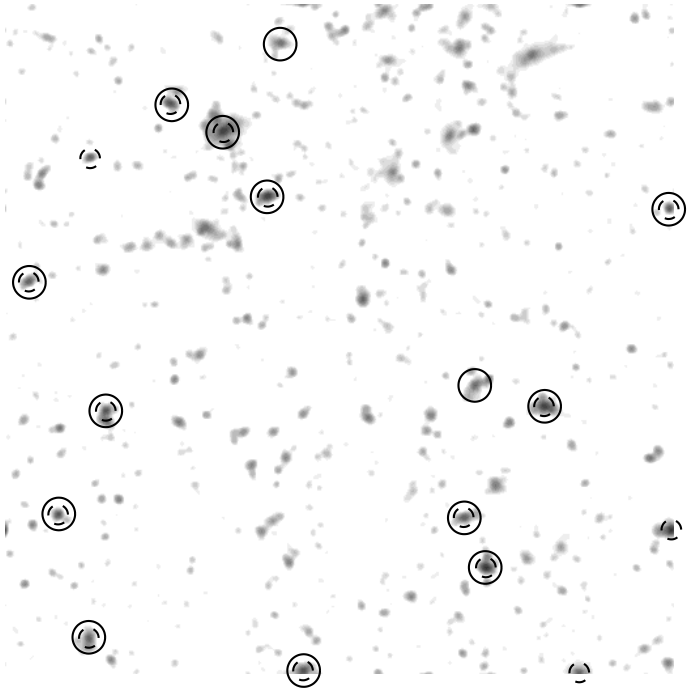}}
\end{center}
\caption{One SZE map, smoothed with a Gaussian of $1'\,$FWHM but with
{\it no noise added}.
The grey-scale is logarithmic, running from $10^{-5.5}$ to $10^{-3.5}$.
There are 14 halos more massive than $4\times 10^{14}h^{-1}M_\odot$ in
the $3^\circ\times 3^\circ$ field (solid circles), of which 12 are found.
There are 15 peaks above a threshold $y_{\rm cut}=5\times 10^{-5}$
(dashed circles).}
\label{fig:example}
\end{figure}

Because the simulation contains no gas we use a semi-analytic model to include
the gas physics.
First we assume that the gas closely traces the dark matter.
This is likely a good approximation in all regions except the innermost
${\cal O}(100)$kpc of the cluster, which for clusters at cosmological
distances will be unresolved by the experiments of interest.
(For an $\Omega_{\rm m}=0.3$ flat cosmology, $100$kpc subtends only $0.26'$
at $z=0.5$ while upcoming survey experiments have beams of $\sim 1'$.)
We ignore the presence of cold gas and stars in the ICM, assuming that the
mass in hot gas is $\Omega_{\rm b}/\Omega_{\rm m}$ of the total.
Second, each cluster is assumed isothermal.
Our assumptions so far are similar to those of
Cooray, Hu \& Tegmark (\cite{CooHuTeg}), but they additionally assumed that
all of the gas in the universe was at fixed temperature, independent of the
mass and virial temperature of the halo in which it resided.  Instead we
assign to each particle in a group a temperature
\begin{equation}
  {k_BT\over {\rm keV}} =
    T_{*} \left( {M\over 10^{15}h^{-1}M_\odot}\right)^{2/3}
          \left( \Delta_c E^2 \right)^{1/3}
\label{eqn:mt}
\end{equation}
where $E(z)\equiv H(z)/H_0$, $\Delta_c$ is the density threshold defining
the mass and $T_{*}$ gives the overall normalization of the relation.
In principle one could use a less steep function for lower masses, since
there is some evidence that the $T-M$ relation becomes shallower at low
mass (Finoguenov, Reiprich \& Bohringer~\cite{FRB}).
However we also expect the gas fraction to drop to lower masses, and
hydrodynamic simulations indicate that we can roughly mimic the effect of
this on our SZE maps by keeping the slope of Eq.~(\ref{eqn:mt}) fixed and
holding $f_{\rm gas}$ at its universal value.
Finally, we smoothly take $T\to 0$ for halo masses below
$10^{13}h^{-1}M_\odot$.  How we do this does not influence the results.
Since most of the SZ emission comes from gas at significant overdensities
(da Silva et al.~\cite{dSBLT}; White, Hernquist \& Springel \cite{WHS})
considering only the particles in the halos when making the maps is a
good approximation.

We use $T_{*}\sim 1$ throughout\footnote{For a power-law spectrum,
$P(k)\sim k^n$, the SZE angular power spectrum scales as
$T_{*}^2\sigma_8^{14/(3+n)}$.  Matching the local temperature function
of rich clusters requires $\sigma_8\sim T_{*}^{-\gamma}$ with
$\gamma\simeq 0.7-0.9$.  Thus increasing $T_{*}$ much above 1 drastically
lowers the predicted SZE fluctuations if one maintains agreement with
the observed XTF.}, which gives good agreement for the redshift evolution
of the mean mass weighted temperature and the angular power spectrum of
$y$ when compared to the results of hydrodynamic simulations
(White, Hernquist \& Springel~\cite{WHS}).
In particular this method provides a better fit to the shape of the simulation
based angular power spectrum, especially over the peak, than semi-analytic
methods (Komatsu \& Kitayama \cite{KomKit}; Cooray \cite{Coo};
Molnar \& Birkinshaw \cite{MolBir}; Holder  \& Carlstrom \cite{HolCar};
Komatsu \& Seljak \cite{KomSel}).
As our approach has isothermal clusters, with a deterministic temperature
derived solely from the mass, this somewhat limits the possible sources of
discrepancy between the semi-analytic and hydrodynamic calculations.
There is a tendency for the N-body results to slightly underpredict (by tens
of percent) the low-$\ell$ tail and to have less low level unresolved emission
than the hydro based maps.
Since the low-$\ell$ tail is sensitive to the volume used in constructing the
maps, and the hydro simulations were run in smaller boxes, we do not regard
this disagreement as significant.

We choose $T_*$ so that the power is close to the level seen by CBI in their
deep field (Mason et al.~\cite{CBI}).  This is a factor of approximately 4
larger than would be predicted by the `concordance' cosmology.
Our results are relatively insensitive to the precise value of $T_{*}$ chosen
or to the treatment of `gas' outside of the virialized regions of halos.

\begin{figure*}
\begin{center}
\resizebox{7in}{!}{\includegraphics{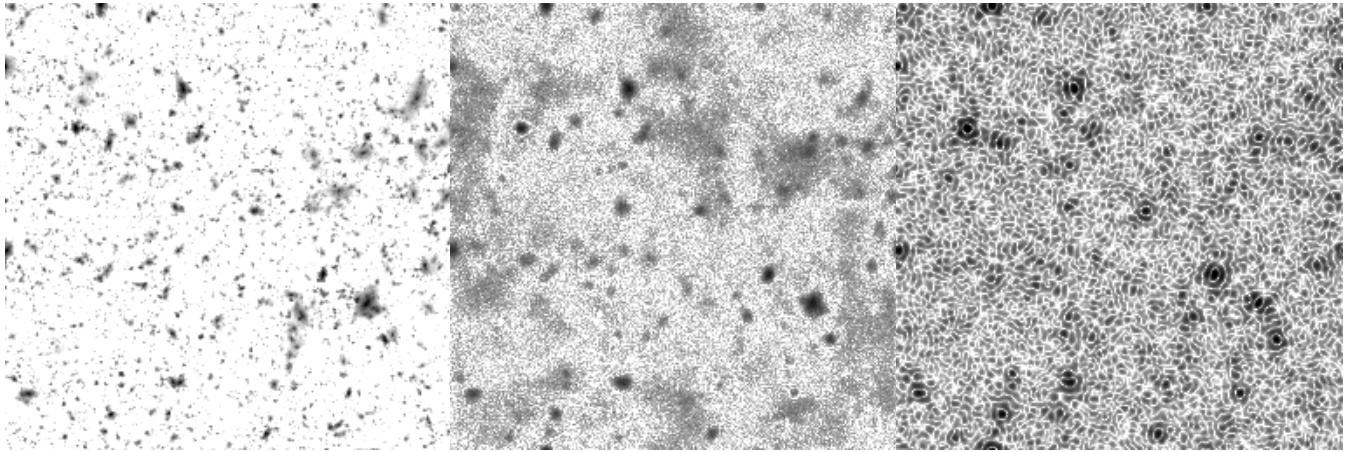}}
\end{center}
\caption{The 3 stages of our mock SZE survey.
(Left) the signal (middle) combined with the primary CMB anisotropies,
smoothed with a $2'\,$FWHM beam and with $5\mu$K per $2'$ pixel of gaussian
random `instrument' noise added (the absolute value is plotted);
(right) the middle map filtered with the matched filter algorithm described
in the text to increase the contrast of the clusters against the background
(the absolute value is plotted).
Each map is $3^\circ\times 3^\circ$ and contains $1024^2$ pixels, rebinned
to $256^2$ for plotting purposes.  The greyscale in each case is logarithmic,
with black 2 orders of magnitude below the peak value (white).}
\label{fig:long}
\end{figure*}

We generate Compton-$y$ maps by integrating for each pixel
\begin{equation}
  y=\int \sigma_{\rm T} n_{e}{k_{\rm B}T_{e} \over m_{e} c^2} dl
  \qquad .
\end{equation}
Here $\sigma_{\rm T}$ is the Thompson scattering cross section, and $n_e$,
$m_e$ and $T_e$ are the electron number density, mass and temperature
respectively.  We assume that within the clusters the gas is fully ionized.
The contribution from each particle is distributed over the pixels with a
spline weighting and a (physical) size equal to the smoothing length of the
simulation.  The temperature fluctuation at frequency $\nu$ is then obtained
{}from the $y$-maps by
\begin{eqnarray}
{\Delta T\over T} &=&
  \phantom{-2}y \left( x{{\rm e}^x+1\over {\rm e}^x-1}-4 \right) \\
  &\simeq& -2y\qquad \mbox{for }\ x\ll 1\, ,
\end{eqnarray}
where $x=h\nu/k_BT_{CMB}\simeq\nu/56.84$GHz is the dimensionless frequency
and the second expression is valid in the Rayleigh-Jeans limit.  In what
follows we shall assume the low-frequency limit unless otherwise stated.

While these assumptions are crude, comparison of the maps with those made
{}from the more sophisticated hydrodynamic simulations of
White, Hernquist \& Springel~(\cite{WHS}) show that they capture many
of the features of the more detailed modeling.  The signals of interest are
dominated by group and cluster sized halos which are quite regular in their
gas properties, allowing a relatively simple minded treatment for our
purposes.
It is important to note that since we are trying to {\it test\/} rather than
{\it calibrate\/} our cluster extraction methods our requirements on the
simulated maps are not too stringent.

We generate 10 random SZE maps, each $3^\circ\times 3^\circ$ with $1024^2$
pixels.
An example map produced in this manner is shown in Figure \ref{fig:example}.
The map is clearly dominated by discrete sources, the galaxy clusters and
rich groups, having a typical size of about $1'$ and a typical amplitude
on the order of mJy.

\subsection{Primary CMB anisotropies}

The primary CMB anisotropies contribute significantly to the SZE signal on
the scales of interest to us, and it is important to consider the effects
of contamination introduced by such fluctuations unless we can use
multi-frequency information to suppress the primary anisotropies.
To investigate this we generate a large CMB map from which we extract a
region of the same size as the SZE map, and add it in as `noise'.
We have used the CMBfast code to generate the CMB power spectrum for our
cosmology (Seljak \& Zaldarriaga~\cite{cmbfast}).
The CMB map is then a random realization of a Gaussian field with this power
spectrum.  We generate random phases in momentum space, and assign amplitudes
to each of the $k$-modes using a distribution whose average value is the
amplitude in the CMB power spectrum.  We have used the flat sky approximation,
in which the $k$-mode in momentum space corresponds to $\ell$ value in the CMB
power spectrum.
The probability distribution function for the amplitudes, $\rho$, of each
$k$-mode is given by
\begin{equation}
  P(\rho^2)={1 \over C_{\ell}} e^{-\rho^2 / C_{\ell}}
\end{equation}
Thus if $\epsilon$ is a random number between 0 and 1, $\rho^2$ will be
given by
\begin{equation}
  \rho^2=-C_{\ell}\,\ln(\epsilon)
\end{equation}

\subsection{Matched filter}

After adding the CMB into the map, we convolve with a Gaussian beam, which
smears the signal and obscures and information on scales much smaller than
the beam size.  To complete the mock observation, we add a few $\mu K$ of
white noise to each pixel corresponding to expected levels of noise from the
electronics used in real observations.
This introduces fluctuations on scales much smaller than the size of the
features in the beam smeared map.

In order to remove the CMB anisotropies, and smooth the small scale noise,
it is necessary to filter the map; we examine a matched filter algorithm to
optimize the contrast of signal to noise.
The specific filter is described in
(Tegmark \& de Oliviera-Costa~\cite{tegfilt}, Eq.~6;
see also Haehnelt \& Tegmark~\cite{HaeTeg}).  It is azimuthally symmetric
and has a radial dependence in Fourier space of
\begin{equation}
  W_{\ell} \propto {1 \over B_{\ell} C^{TOT}_{\ell}}
  \qquad .
\label{eqn:filter}
\end{equation}
Here $B_{\ell}=e^{-\theta^2 \ell(\ell+1)/2}$ and $C^{TOT}_{\ell}$ refers to
the sum of the power spectra of elements to be removed from the map, in our
case the CMB and white noise.
The noise power spectrum is given by
$C^{N}_{\ell}=(\sigma_{\rm pix} \theta_{\rm pix})^{-2} B_{\ell}^{-2}$
(e.g.~Tegmark \& Efstathiou~\cite{TegEfs}).
For the normalization of the filter, we use the $1\sigma$ value of the noise
(only) map after it has been filtered with $W_{\ell}=1/B_{\ell}C^{TOT}_{\ell}$.
In this way, we can compare mock observations with different beam sizes and
levels of noise in terms of the statistical significance of the cluster SZE
signal above the white noise in the map.
Note that in some cases the instrumental noise level will be significantly
below the `noise' induced from CMB anisotropies, so amplitudes in these
maps can be many `noise sigma'.

The matched filter formally maximizes the efficiency with which the SZE
survey will locate isolated clusters because it maximizes the contrast
between the signal and the background noise.
In a real cluster survey, however, the analysis team may wish to sacrifice
efficiency somewhat if in doing so it can provide a substantial gain in the
completeness of the survey.
In many cases, particularly when the beam size is large, the filter in
Fourier space is very narrow, and filters out much of the signal along
with the noise.  In addition, a narrow filter in Fourier space develops
significant side lobes in real space.
When such a filter is convolved with the density spike at the location
of the cluster, it enhances both the central peak and an annular region
at the location of each side lobe.
In the most extreme cases, several concentric rings may appear around the
central maximum that marks the location of the cluster.  These rings can
overlap with other rings from nearby structures in complicated ways.
If these rings are misidentified as separate clusters, they negatively
impact the efficiency\footnote{Such considerations will obviously apply
to any filtering technique with a compensated filter, such as for example
a wavelet based approach.  The key parameter is the mean separation of bright
sources compared to the filtering scale.  If the separation is smaller than
the filtering scale, the compensated filter needs to be used with caution.}.
Thus making the filter a little wider than the matched filter for experiments
with lower resolution helps by improving the completeness significantly while
it ameliorates the ringing.

\subsection{Peak finding}

Even a glance at Figures \ref{fig:example} and \ref{fig:long} is enough to
show that finding peaks in an SZE observation is fraught with difficulties.
The peaks are often irregularly shaped, contain significant substructure and
merge with neighboring peaks.
We spent considerable time trying different methods of detecting substructure,
merging peaks and imposing thresholds on either total flux or peak flux.
We found that exactly which peaks passed which cuts depended on how these
issues were handled, but were unable to find a strategy which worked in every
case.
Smoothing the maps with a resolution of $1'$ or more, matched to the angular
size of clusters at cosmological distances, mitigated some of the sensitivity
but did not entirely eliminate the dependence on peak finding properties.
In particular, which peaks are found in dense environments (which may for
example correspond to interacting halos) depends on ones choice for peak
edges and merger criteria.  This is an analogue of the problem we found in
defining clusters in the 3D dataset under similar conditions.

\begin{figure}
\begin{center}
\resizebox{3.5in}{!}{\includegraphics{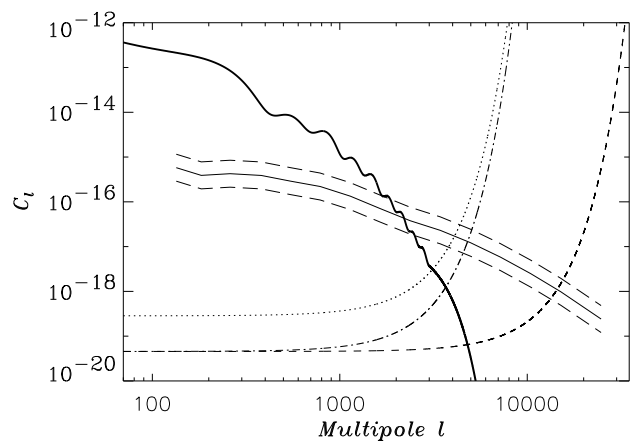}}
\end{center}
\caption{The role of sensitivity and resolution in SZE surveys.
The three types of curves, from top to bottom at left of plot) are the
power spectra of the primary CMB anisotropies (thick solid),
the SZE signal (an average of the power spectra in our ten maps;
solid+dashed band),
and the instrument noise (dotted, dot-dashed and dashed).
The dotted curve has a $3'\,$FWHM Gaussian beam with $4\mu$K of noise
per $1'$.  The dot-dashed curve has the same beam size, but $2\mu$K noise
per $1'$.  The dashed curve has a $1'$ beam with $2 \mu K$ of noise.}
\label{fig:p1}
\end{figure}

However we did notice that while the exact numbers were sensitive to the
peak finder, the general trends we find (below) were not very sensitive.
For this reason we chose the simplest peak finder, with the fewest adjustable
parameters, in order to present our results.
Specifically we used a simple algorithm, similar in nature to the one in
White, van Waerbeke \& Mackey~(\cite{WhivWaMac}).
We first record and number all the pixel locations of local maxima in the map.
We search around each local maximum and include in the extended peak all
pixels with a value greater than 25\% of the maximum value, out to a maximum
radius of 10 pixels.
All peaks are extended at the same rate, so that adjacent peaks do not merge
into one object.  This has a tendency to split objects with significant
substructure into multiple peaks, but for smoothed maps such a situation is
reasonably rare.
The algorithm returns a peak number (or no peak) for every pixel in the map.
We have tested sensitivity to the 25\% criterion for associating a pixel with
a peak and find that, for the smoothed maps, completeness and efficiency are
unaffected by moderate changes in this parameter.
The `value' associated with a peak is the peak temperature fluctuation.
Since we are smoothing on scales comparable to the total size of a peak
it makes little difference whether we choose peak temperature or total
fluctuation.
The fraction of peaks in the mock observation that matched at least one halo
is the efficiency, and the fraction of halos that matched a peak is the
completeness.

Certainly more complicated peak finding methods could be employed, a
different filter could be tried (e.g.~a wavelet based method such as
described in Cay\'on et al.~\cite{Cay} or Herranz et al.~\cite{HSHBDML},
which corresponds to bandpass filtering)
and more sophisticated modeling (e.g.~along the lines of the CLEAN
algorithm: H\"ogbom \cite{Hog}; Clark \cite{Cla}) could remove some of the
artifacts introduced by the filtering.
Because of this we feel that improved strategies for finding peaks in such
maps is an area worthy of more attention.

\section{Results}

We find that the number of clusters that can be recovered from the mock SZE
survey is a strong function of the beam size, the level of noise, and the
threshold at which we identify cluster candidates.
The level of contamination of the cluster candidates also depends strongly
on these quantities.
In general, decreasing the beam size, decreasing the noise, and decreasing
the threshold for identifying candidates in the SZE map will improve
completeness at the expense of efficiency.
Since efficiency is not the exact inverse of completeness, some combinations
of these three parameters will yield better results than others.  
We also find that multi-frequency information can be extremely valuable in
finding clusters, by removing the dominant noise source at large angular
scales.

Figure \ref{fig:example} shows a `perfect' observation by a multi-frequency
instrument.  The Compton-$y$ map has been smoothed at $1'$, but no noise
or CMB has been added.  The grey-scale has been selected to emphasize the
most prominent peaks.  We have indicated the peaks corresponding to the
most massive clusters in the field by solid circles, and the biggest peaks
in the field by dashed circles.  As one can see most of the massive clusters
are easily recovered with few false positives, but there is not a 1-1
correspondence even at these high thresholds.
The source of the disagreement can be traced to the large number of objects
in the maps (confusion) and a scatter in the relation between the mass and
the observable SZE (White, Hernquist \& Springel~\cite{WHS}).
The sources of this scatter are threefold:
First, there is a time evolution in the relation between the mass of a cluster
and its temperature, Eq.~(\ref{eqn:mt}).
Since the clusters are from a large range in redshifts, this causes some
scatter.
Secondly, projection effects are non-negligible.
In fact, clusters have a tendency to form in over-dense regions, often at the
intersection of a beaded filamentary structure, increasing the probability
of non-trivial line-of-sight projection.
Finally, clusters are not spherical and the signal strength depends on their
orientation.
Since the lower mass halos are intrinsically more numerous, even
misidentifying a small fraction of them can negatively impact the
survey efficiency.

Of course the situation depicted in Figure \ref{fig:example} is highly
unrealistic.  To truly asses how well a survey can find clusters we need
to include both astrophysical and instrument noise.
The images in Figure \ref{fig:long} display the various stages of our mock SZE
survey.  The pure SZE map is shown in panel one.
Panel two shows how the SZE signal from clusters is clouded by adding the CMB,
convolving with a Gaussian beam of $1'$ angular extent, and further adding
Gaussian random noise.  Panel 3 shows the map in panel two after it has been
filtered with the matched filter described above.  

Figure \ref{fig:p1} illustrates the role of the size of the beam and the
noise level in locating clusters with an SZE survey.
The three types of curves are the power spectra of the primary CMB
anisotropies, the SZE signal (an average of the power spectra in our ten maps),
and the instrument noise.
(In the case of multi-frequency observations the primary CMB signal could
be eliminated or at least strongly suppressed.)
The range of scales over which an experiment is sensitive to the SZE
are those scales for which the SZE signal is not overpowered by either the
CMB or the instrument noise.
The dashed lines above and below the SZE power spectrum are meant to remind
the reader that the normalization of the power spectrum depends non-trivially
on the cosmology we have simulated, particularly on the value of $\sigma_8$,
and could change by a factor of 2 or 3 with different initial assumptions.
In circumstances where the window of SZE sensitivity is narrow, such a change
profoundly affects the projected yield of an experiment.
The three noise curves are plots of  $C_{\ell}^N$, which depends on the level
of noise and on the beam size.  For clarity, all noise levels reported are
in $\mu K$ per $1'$ pixel, regardless of the beam size.
The dotted curve has a $3'$ Gaussian beam at full width half maximum,
with $4 \mu K$ of noise.  The dot-dashed curve has the same beam size,
but the level of noise has been reduced to $2 \mu K$.
It is clear that reducing the level of noise has a relatively small impact
on the range of scales that can be probed.  In contrast, the dashed curve
which has a beam size of $1'$ with $2 \mu K$ of noise, demonstrates that
decreasing the beam size has a dramatic effect on the range of sensitivity.
Alternatively, using multi-frequency observations to reduce the contribution
of the primary anisotropies can allow one to work at lower $\ell$.

\begin{figure}
\begin{center}
\resizebox{3.5in}{!}{\includegraphics{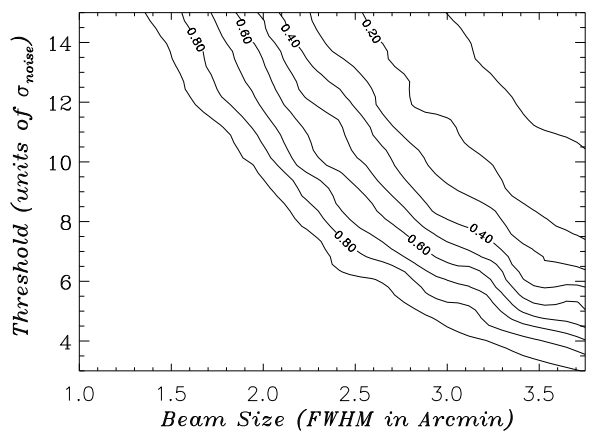}}
\end{center}
\caption{Completeness in finding halos more massive than
$3\times 10^{14}\,h^{-1}M_\odot$ in maps with noise $5\mu$K per $1'$ pixel
with peaks selected as local maxima.}
\label{fig:p2}
\end{figure}

\begin{figure}
\begin{center}
\resizebox{3.5in}{!}{\includegraphics{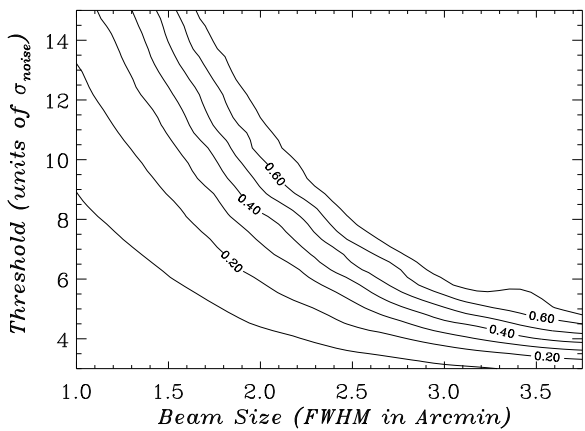}}
\end{center}
\caption{Efficiency at finding halos more massive than
$3\times 10^{14}\,h^{-1}M_\odot$ in maps with noise $5\mu$K per $1'$ pixel
with peaks selected as local maxima.}
\label{fig:p3}
\end{figure}

\subsection{Multi-frequency observations}

We begin by examining maps which do not include the contribution from the
CMB.  Such maps could be obtained using multi-frequency observations (we will
not address here the methods by which component separation is done, for a
recent survey of the literature see e.g.~Vielva et al.~\cite{VBHMLST} or
Herranz et al.~\cite{HSHBDML}) of the same field.
This removes the largest source of confusion from the SZE maps.
To generate these maps we converted the Compton-$y$ maps to temperature
fluctuation maps, smoothed them with a Gaussian filter using an FFT, added
an appropriate level of pixel (white) noise and once again smoothed the maps.
This last stage of smoothing was necessary since our pixel scale is
significantly smaller than the beam, leading to a large per pixel noise.
Smoothing the maps reduces this noise with little effect on the signal.

We find that for noise levels as low as $5\mu$K per $1'$ pixel both the
completeness and efficiency can be very good at high angular resolution.
For thresholds above $5\sigma$ more than 80\% of the peaks correspond
to rich clusters above $3\times 10^{14}h^{-1}M_\odot$, and for beams of
$1'\,$FWHM such a cut recovers 75\% of the existing clusters above this
mass threshold.  As long as a cut of at least $4\sigma$ is applied the
efficiency is greater than 60\%.  For a beam smaller than $2'\,$FWHM more than
half of the clusters can be recovered.
However such a low noise level may be optimistic for upcoming experiments.
If we increase the noise to $10\mu$K per $1'$ pixel the situation degrades.
The best completeness is now around 60\% and only at low thresholds, for
which the efficiencies are low.  To avoid confusion the beam also needs to
be less than $1.5'\,$FWHM.

As a concrete example we consider an idealization of the {\sl Planck\/}
satellite surveying $\sim 70\%$ of the sky.
In terms of cluster finding, {\sl Planck\/}'s capabilities are primarily
limited by its resolution, many distant clusters subtending a much smaller
angle that {\sl Planck\/}'s beam size.  
As such, the sample found by {\sl Planck\/} will be biased in that it will
tend to identify the nearest clusters
(see Aghanim et al.~\cite{Aghanim} or Kay, Liddle \& Thomas \cite{KayLidTho}
for more details and Herranz et al.~\cite{HSHBDML} for a recent study of the
expected {\sl Planck\/} SZE catalog).
We assume that component separation has left no primary CMB signal in our
SZE map.  In particular, we optimistically computed the completeness and
efficiency with equal parts noise from the 217 GHz (no SZE) and 353 GHz
channels, added in quadrature.
These channels both have an angular resolution of $5'$.
Within this (overly) simplistic approximation {\sl Planck\/} finds close to
half of the objects more massive than $3\times 10^{14}\,h^{-1}M_{\odot}$,
but with our simple peak finding algorithm the efficiency is below 20\%.
Over $70\%$ of the sky this is a very large sample, useful for many studies
of clusters.  However a more sophisticated cluster finding algorithm would
need to be employed before this sample could be used as a cosmological probe.
Improved methods for finding massive clusters in low resolution but
multi-frequency data is a topic which we intend to pursue further in a
future publication.

\begin{figure}
\begin{center}
\resizebox{3.5in}{!}{\includegraphics{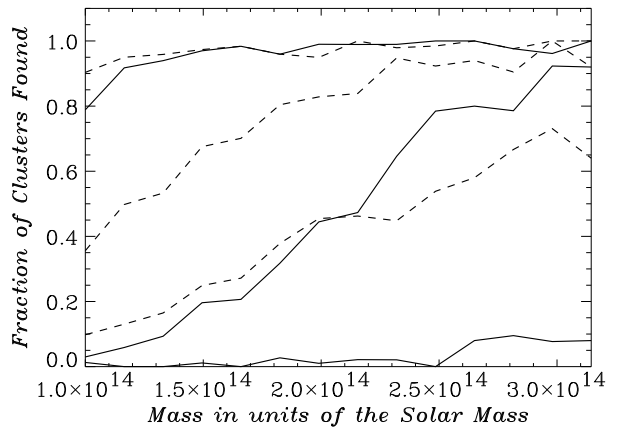}}
\end{center}
\caption{The fraction of the total number of clusters in each mass bin of 
width $1.75\times 10^{13}h^{-1}M_{\odot}$ found in mock surveys of three
different beam sizes: $1'$, $2.5'$ and $4'$ (top to bottom).  
The solid curves indicate the results when the matched filter is used,
the dashed curves show the improvement in completeness when a less severe
filter is employed.  Slope in the curves indicates a selection bias.}
\label{fig:mhist}
\end{figure}

\subsection{Spatial filtering}

Let us now turn to the case of single frequency maps.  In this situation we
must remove the large CMB contribution by using its spatial structure.
Figures \ref{fig:p2} and \ref{fig:p3} are contour plots of the completeness
and efficiency that can be achieved by a single frequency experiment, using
the optimal filter of Eq.~(\ref{eqn:filter}), for various beam sizes.
Again we consider all rich clusters above $3\times 10^{14}h^{-1}M_\odot$.
The contours are computed at a constant noise level of $5\mu$K per $1'$ pixel.
On the $y$ axis is the threshold used on the filtered plot to identify
cluster candidates.  The threshold indicates the statistical significance
of the cluster candidate above the level of instrument noise.
The $1\sigma$ value is determined by filtering the noise map without the
SZE or CMB signal, and computing the resulting variance.
In Figure \ref{fig:p2} the completeness improves as the resolution gets
better (smaller beam), however some percentage of the clusters are
overlooked as the candidate threshold is raised.
Raising the threshold is useful however in terms of improving the efficiency
of the survey, as can be seen in Figure \ref{fig:p3}.
Because all of the candidates need to be followed up for redshift information
(and positive identification as a real cluster, since the contamination for
some experiments can be large), efficiency is a high priority.
Since the contours in Figures \ref{fig:p2} and \ref{fig:p3} are not
completely parallel, there is not a simple trade off between completeness
and efficiency, but rather there are regions that are clearly somewhat better
compromises than others.

\begin{figure}
\begin{center}
\resizebox{3.5in}{!}{\includegraphics{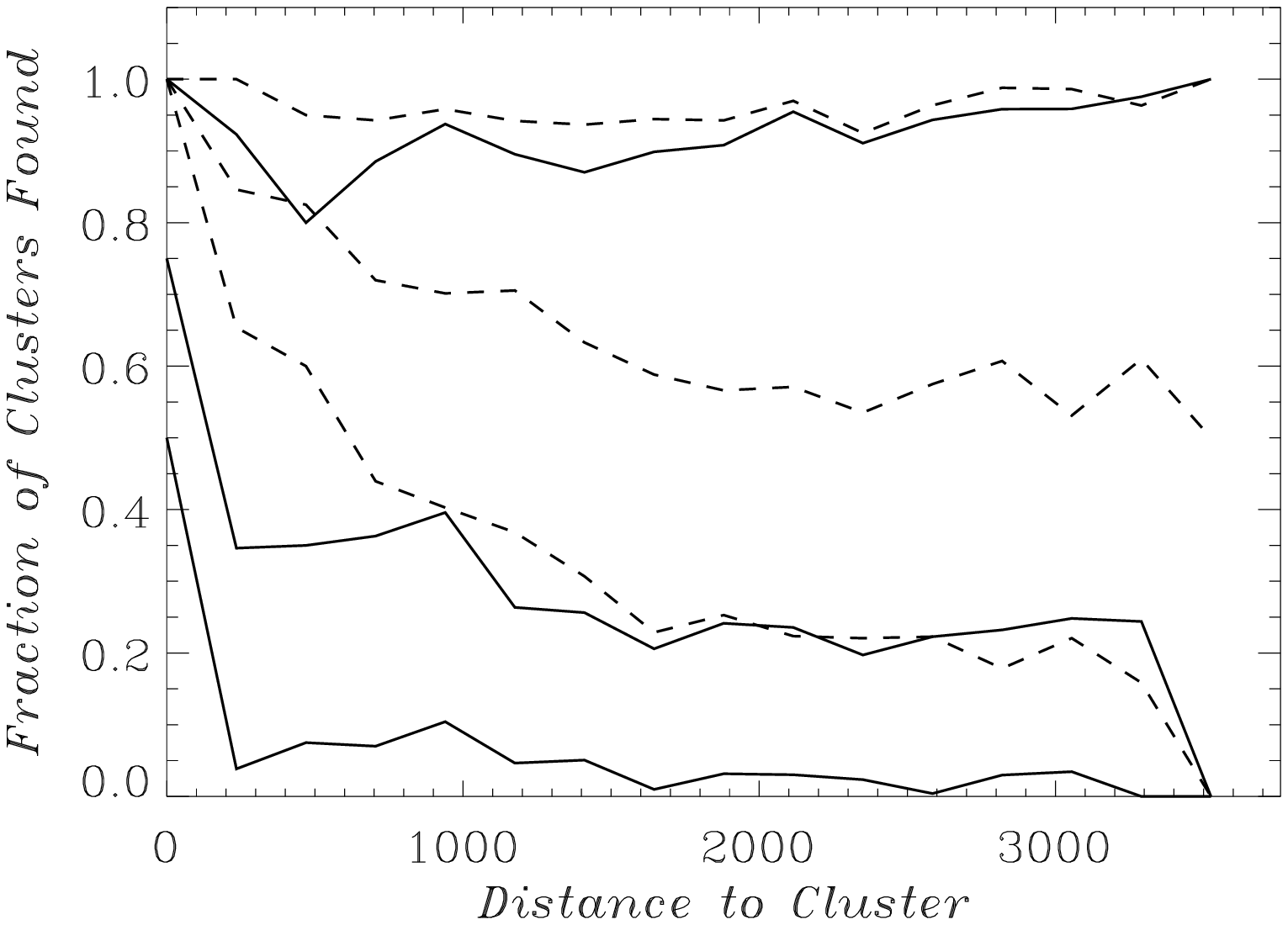}}
\end{center}
\caption{The fraction of the total number of clusters in each distance bin of 
width $235h^{-1}$Mpc found in mock surveys of three different beam sizes:
$1'$, $2.5'$ and $4'$ (top to bottom).  
The solid curves indicate the results when the matched filter is used,
the dashed curves show the improvement in completeness when a less severe
filter is employed.  Slope in the curves indicates a selection bias.
In this cosmology $z=1$ is at (approximately) $2300\,h^{-1}$Mpc.
The distribution of halos more massive than $3\times 10^{14}\,h^{-1}M_\odot$
peaks at $z\sim 0.6$ or $1400\,h^{-1}$Mpc and falls by two orders of
magnitude by $z=2$ ($3600\,h^{-1}$Mpc).}
\label{fig:dhist}
\end{figure}

The very low effiencies evident in the bottom left of Figure \ref{fig:p3}
should not be read as confusion by noise in the map.
There are no noise peaks of $10\sigma$ in these maps, even including the CMB!
Rather they come from lower mass halos which happen to give rise to peaks
which are many sigma above the noise when the beam is small.  For a $1'\,$FWHM
beam at $5\mu$K even halos with $10^{14}\,h^{-1}M_\odot$ can give rise to an
$>5\sigma$ peak, and these outnumber the $M>3\times 10^{14}\,h^{-1}M_\odot$
halos 10:1.
Raising the threshold even further (off the top of the plot) removes the
lower mass peaks and improves efficiencies, but at the expense of completeness
as in the above examples.

\subsection{Non-optimal filter}

Another degree of freedom is to change the shape of the filter instead of
adjusting the threshold.  As mentioned earlier, broadening the filter can
improve completeness without clobbering the efficiency because the increase
in noise confusion is moderately compensated by the decrease in ringing.
Broadening the filter can also decrease both the mass and distance biases
that occur in SZE surveys with limited angular resolution.
Figures \ref{fig:mhist} and \ref{fig:dhist} show the completeness for three
different beam sizes, binned in mass and distance bins respectively.
These plots demonstrate that while high resolution experiments are relatively
unbiased, surveys with a large beam size will tend to deselect distant or
less massive clusters because the angle they subtend is smaller than the
resolution of the experiment.
By broadening the filter
(dashed lines in Figs.~\ref{fig:mhist} and \ref{fig:dhist}),
the completeness of the sample improves, as does the bias in some cases.
Decreasing the bias, or at very least quantifying it, is critical if the
survey is to be used to constrain the cosmological parameters.

We find that broadening the filter on the high $\ell$ side (noise side)
in Fourier space is ineffective because the increase in the noise confusion
is too great, completely destroying the efficiency.
Instead we have filtered less harshly on the low $\ell$ portion of the mock
survey, replacing the $1/C_{\ell}^{CMB}$ portion of the filter with a much
wider half Gaussian that peaks at the same value of $\ell$.
The improvement in the completeness is naturally accompanied by a corresponding
decrease in the efficiency.
The efficiencies for $4'$, $2.5'$,and $1'$ beam sizes in the matched filter
case were 93\%, 86\%, and 52\% respectively.
In the modified filters, the width of the Gaussian was chosen so that the
efficiency would be approximately 45\%.
These widths were 1600, 2500, and 2500 $\ell$ values respectively. 
This method could be used to satisfy any efficiency requirement lower than
the efficiency of the matched filter, in order to improve completeness.
While we did not explicitly try it we expect that the mexican hat
wavelet, which corresponds to a filter $x^2e^{-x^2}$ with $x\propto\ell$,
would also work well.

\section{Conclusions}

The SZE offers a new and potentially very powerful method for finding high
redshift clusters of galaxies.  Because the amplitude of the effect is
independent of the distance to the cluster, it appears to be one of the best
techniques for constructing a large sample of high-$z$ clusters.

Finding clusters with an SZE survey is however fraught with complications,
some of which we have begun to address here using mock observations of
simulated maps.  While there remains significant uncertainty in the overall
level of the SZE angular power spectrum and our modeling of the effect has
been somewhat crude, we already see that the requirements on frequency
coverage, angular resolution and noise are quite severe for experiments
hoping to find large samples of clusters through the SZE.
In single frequency maps, the primary CMB anisotropies prove to be a large
contaminant.  Indeed for experiments with angular resolutions of $>1'$ there
is little spatial range where the SZE signal dominates.

Many important effects must be balanced when designing the experiment and
analyzing the data.  The angular resolution of the instrument used is of
paramount importance, a key element in determining the yield of the survey.
Decreasing the beam size improves the completeness, and more importantly 
decreases the bias against distant and lower mass clusters.
For any given resolution, however, there are adjustable parameters in the
data analysis that can help reduce the bias and maximize completeness, at
the moderate expense of efficiency.
Efficiency is still a vital part of the survey design, however, because 
each cluster candidate must be followed up for positive identification and
redshift information.
The intrinsic contamination that occurs as a result of projection effects
makes a follow up required, even if the precise redshift is not needed.  
To obtain a high level of completeness with correspondingly high efficiency
requires a multi-frequency observation with angular resolution of $1'$ or
better, and noise at or below $10\mu$K per $1'$ pixel.

We have found that aggressive spatial filtering, to enhance the clusters
against the background, can have the unwanted side effect of introducing
ringing into the maps.  Given the large number of sources in a typical
simulated map, overlapping `rings' can produce significant false detections.
Multi-frequency information could help reduce some of the pitfalls inherent in
a single frequency analysis, and may provide a less severe alternative than
matched filtering to disentangle SZE clusters from primary CMB anisotropies.
For any of these approaches, more sophisticated methods of identifying peaks
in the SZE map (e.g.~matched filtering) need to be investigated.
Better data analysis offers the hope of increased completeness without a
sacrifice in efficiency.

\bigskip
The authors would like to thank Nils Halverson, Erik Reese and Chris Vale
for many useful discussions on this work.
The simulations used here were performed on the IBM-SP2 at the National
Energy Research Scientific Computing Center.
This research was supported by the NSF and NASA.  M.W. was supported by
a Sloan Foundation Fellowship.

\end{document}